\documentclass[reprint,superscriptaddress,amsmath,amssymb,aps]{revtex4-2}
\usepackage{graphicx,color,hyperref,float,amsmath,chemformula,comment}
\begin{document}
\title{Thermoelectrics properties of two-dimensional materials with combination
of linear and nonlinear band structures}
\author{Andri Darmawan}\email{akholildarmawan@stu.kau.edu.sa}
\affiliation{Department of Physics, King Abdulaziz University, Jeddah 21589, Saudi Arabia}
\author{Edi Suprayoga}\email{edi.suprayoga@brin.go.id}
\affiliation{Research Center for Quantum Physics, National Research and 
  Innovation Agency (BRIN), South Tangerang 15314, Indonesia}
\author{Ahmad~R.~T.~Nugraha}\email{ahmad.ridwan.tresna.nugraha@brin.go.id}
\affiliation{Research Center for Quantum Physics, National Research and 
  Innovation Agency (BRIN), South Tangerang 15314, Indonesia}
\author{Abdullah~A.~AlShaikhi}
\affiliation{Department of Physics, King Abdulaziz University, Jeddah 21589, Saudi Arabia}
\affiliation{Physical Science and Engineering Division, King Abdullah University of Science and Technology (KAUST), Thuwal 23955-6900, Saudi Arabia}
\begin{abstract}
We investigate thermoelectric (TE) properties of two-dimensional materials possessing two Dirac bands (a Dirac band) and a nonlinear band within the three-(two-)band model using linearized Boltzmann transport theory and relaxation time approximation. In the three-band model, we find that combinations of Dirac bands with a heavy nonlinear band, either a parabolic or a pudding-mold band, does not give much difference in their TE performance. The apparent difference only occurs in the position of the nonlinear band that leads to the maximum figure of merit ($ZT$). The optimum $ZT$ of the three-band model consisting of a nonlinear band is found when the nonlinear band intersects the Dirac bands near the Fermi level. By removing the linear conduction band, or, in other words, transforming the three-band model to the two-band model, we find better TE performance in the two-band model than in the three-band model, i.e., in terms of higher $ZT$ values. 
\end{abstract}
\date{\today}
\maketitle
\section{Introduction}
Heat is often considered an energy waste, yet it can be converted into useful electrical energy using thermoelectric (TE) materials. However, compared to other energy conversion schemes, TE devices often have lower efficiency. One can assess the TE performance through the so-called figure of merit: $ZT = \frac{S^2\sigma}{\kappa}T$, where $S$ is the Seebeck coefficient, $\sigma$ is the electrical conductivity, $T$ is the operating temperature, and $\kappa$ is the total thermal conductivity. The total thermal conductivity, $\kappa$, is a sum of the electronic thermal conductivity ($\kappa_e$) and the lattice thermal conductivity ($\kappa_{ph}$), i.e., $\kappa = \kappa_e+\kappa_{ph}$.  The figure of merit is also proportional to the power factor (PF) by the following relation: $ZT = \mathrm{PF}. T / \kappa$, where $\mathrm{PF} = S^2 \sigma$.  From the expression of ZT, a good TE material should possess good electrical conduction and thermal isolation.  In other words, we should maximize the Seebeck coefficient and electrical conductivity, but, at the same time, minimize the thermal conductivity to obtain a good TE material. Unfortunately, the interrelation between those parameters makes it difficult to find materials with high ZT~\cite{majumdar2004thermoelectricity,vining2009inconvenient,snyder2011complex,sootsman2009new,zeier2016thinking,zebarjadi2012perspectives,zhang2015thermoelectric,zhu2017compromise,mao2018advances,he2017advances,yang2016tuning,gorai2017computationally,jain2016computational}. For example, based on the Wiedemann-Franz law, the ratio between electrical conductivity $\sigma$ and electrical thermal conductivity $\kappa_e$ is constant. Therefore, it is tricky to get material with high $\sigma$ simultaneously with low thermal conductivity $\kappa_e$~\cite{zhao2014ultralow,takahashi2012thermoelectric}.

To solve the TE trade-off problem, there have been some promising proposals, such as optimizing carrier concentration~\cite{CHEN2012535,Zhou2016}, using low-dimensional materials~\cite{hicks1993effect,hung2016quantum,dresselhaus2007new,urban2007synergism,lee2016thermal,eslamian2017inorganic,bottner2006aspects}, utilizing band energy convergence~\cite{pei2011convergence,zeier2014band,zhang2014high,gayner2016recent}, and considering hierarchical architecture~\cite{biswas2012high}. Two-dimensional (2D) materials such as 2D Dirac materials have been shown to feature good electronic and thermal properties which are potential for TE materials~\cite{hippalgaonkar2017high,duan2016high}. Moreover, a solely parabolic band in a 2D material has been found to give more desirable TE properties than a three-dimensional (3D) material due to the quantum confinement that can lead to an improvement in the figure of merit~\cite{hicks1993effect}.  

The band engineering method such as tuning the band gap~\cite{pei2011alloying} or the effective mass~\cite{bilc2004resonant,heremans2012resonant} in terms of the curvature of the band also becomes an effective way to scan potential TE materials because this method uses a relatively cheap computational method by considering only the energy dispersion relation $E(k)$ and the scattering lifetime $\tau(E)$.  One can tune the bandgap or change the combination of the band structures to know the optimized structure that will give better TE properties.  Several works related to the band engineering have been performed for many types of band structures, such as pudding-mold bands~\cite{usui2017enhanced,usui2013large,kuroki2007pudding,isaacs2019remarkable,wei2020strain}, parabolic bands~\cite{xia2019leveraging,xia2019high,adhidewata2021thermoelectric}, and linear Dirac bands~\cite{hasdeo2019optimal}.  In particular, the existence of a heavy band alongside the Dirac bands gives a high power factor (PF) in a semimetal like \ch{CoSi}.  This result is contrary to the mainstream impression that semimetals are generally poor for thermoelectricity~\cite{xia2019high}.  Furthermore, the existence of a heavy parabolic band that intersects Dirac bands at the Fermi level is found to enhance PF due to the improved electron-phonon scattering in the form of a sharp spike of density of states (DOS)~\cite{xia2019leveraging}. Specifically, the existing heavy band acts as a filter for the low-energy carriers to be excited. This unbalance condition will boost the Seebeck coefficient.  However, rather than the heavy parabolic band, a single pudding-mold band can increase the electrical conductivity resulting in a higher power factor and figure of merit~\cite{usui2017enhanced}.  Regarding this fact, we are intrigued to ask, can the pudding-mold band alongside the Dirac bands lead to a better TE performance than the parabolic one? 

To answer the question, in this paper, we will discuss the TE properties of 2D materials that possess Dirac bands with a heavy band. As we will see later, the existence of a heavy band alongside Dirac bands will improve the thermoelectric performance even with a narrow bandgap. We achieved the maximum value of $ZT$ when the heavy band intersects the Dirac bands at the Fermi level for the parabolic band but not for the other heavy bands~\cite{xia2019leveraging}. However, removing one linear band will give a better figure of merit $ZT$ for every type of heavy band.
\section{Model and methods}
\label{sec:th}
\begin{figure}[tb]
    \centering
    \includegraphics[width=8cm]{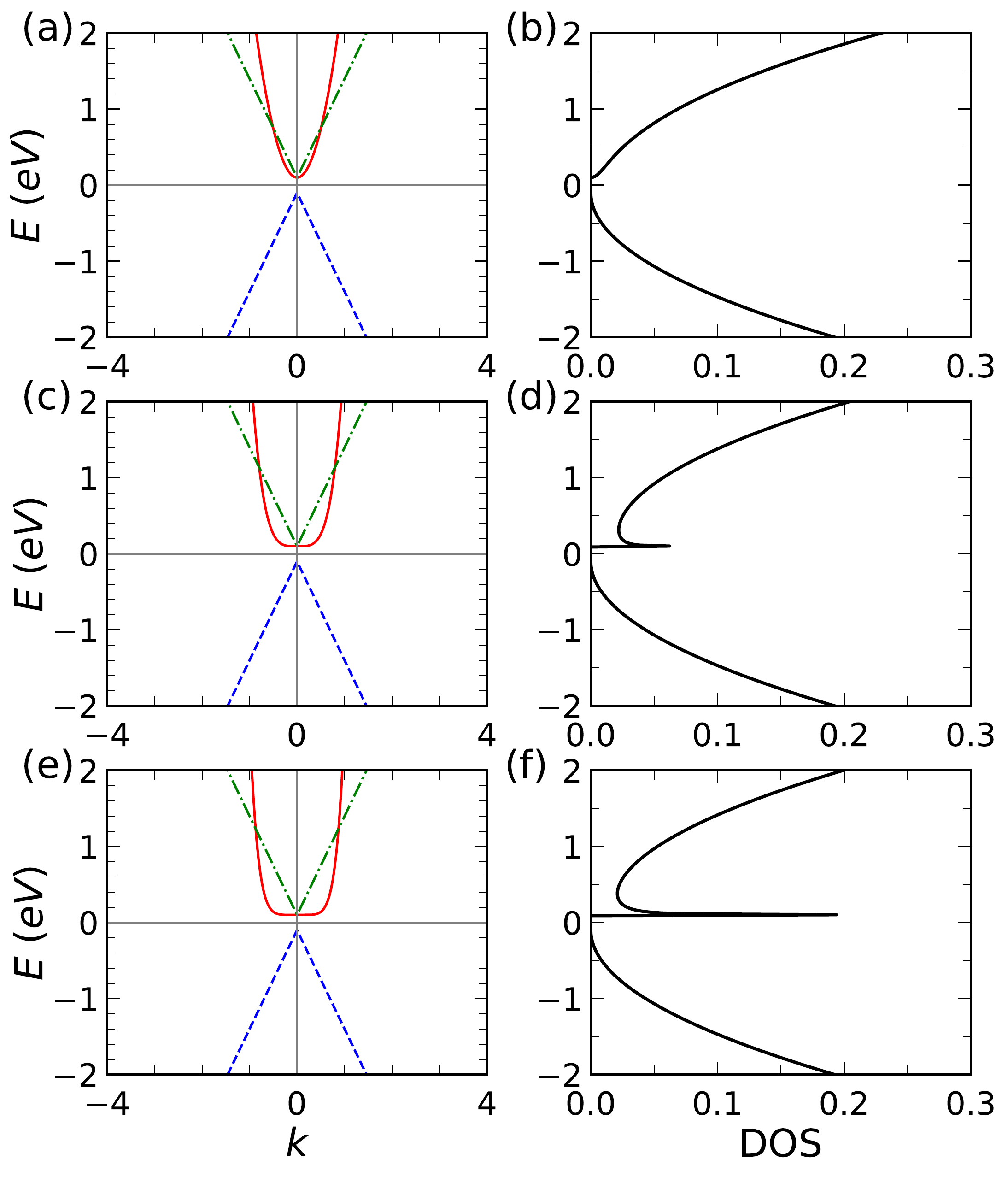}
    \caption{Schematics of the energy dispersion relations and density of states for three-band systems that consist of a parabolic band in (a) and (b), 4th-order pudding-mold band in (c) and (d), and 6th-order pudding-mold band in (e) and (f).}
    \label{bandDOS3}
\end{figure}

\begin{figure}[h]
    \centering
    \includegraphics[width=8cm]{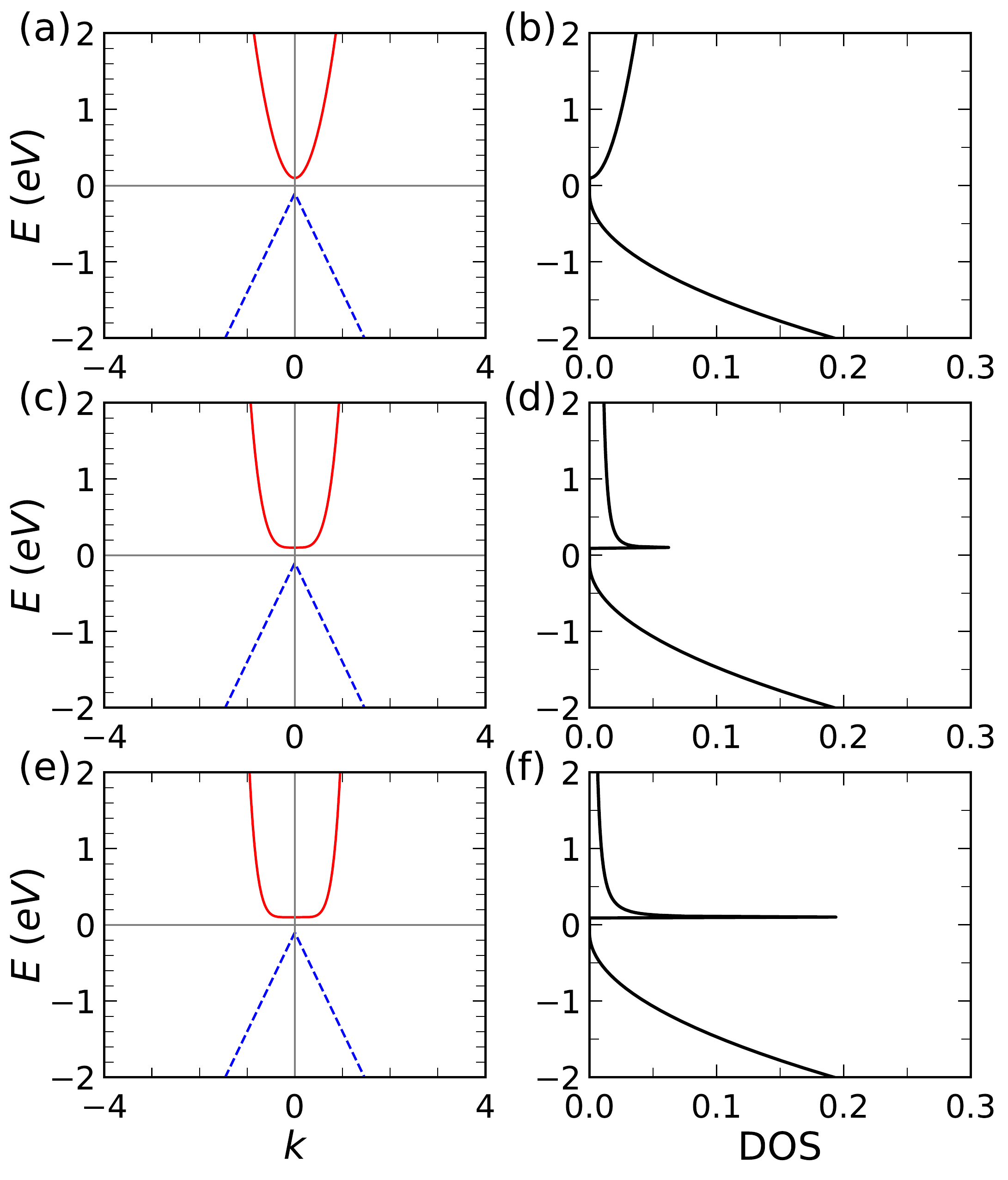}
    \caption{The energy dispersion and density of states for two-band system hat consist parabolic band in (a) and (b), pudding-mold 4th order band in (c) and (d), and pudding mold 6th order band in (e) and (f).}
    \label{bandDOS2}
\end{figure}

We assume the electronic energy dispersion $E(\mathbf{k})$ for the linear/Dirac and nonlinear/heavy bands by a common formula:
\begin{equation}
    E^{(n)} = A_n|\mathbf{k}|^n+\Delta_n,
    \label{eq:disp}
\end{equation}
where $n$ is the order of dispersion, $\mathbf{k}$ is the electronic wavevector, $\Delta$ is the energy shift from Fermi level, and $A_n$ is the normalization constant. The order $n=1$ and $n=2$ correspond to the Dirac and parabolic bands, respectively, while $n=4$ and $n=6$ both correspond to the pudding-mold band. The three-band system with parabolic, pudding-mold 4th order, and pudding-mold 6th order are depicted in Figs.~\ref{bandDOS3}(a), \ref{bandDOS3}(c), and \ref{bandDOS3}(e) respectively with the corresponding DOS in Figs.~\ref{bandDOS3}(b), \ref{bandDOS3}(d), and \ref{bandDOS3}(f). By removing the linear conduction band, the two-band system is modeled as shown in Figs.~\ref{bandDOS2}(a), \ref{bandDOS2}(c), and \ref{bandDOS2}(e) with the corresponding DOS in Figs.~\ref{bandDOS2}(b), \ref{bandDOS2}(d), and \ref{bandDOS2}(f).  

We use the Boltzmann transport theory within linear response regime and apply relaxation time approximation (RTA). Using this approximation, thermoelectric properties can be calculated from transport coefficients with the following TE kernel integration:
\begin{equation}
    \mathcal{L}_i = \int_{-\infty}^{\infty} \mathcal{T}(E)(E-\mu)^i\left(-\frac{\partial f}{\partial E}\right) dE,
\end{equation} 
where $\mathcal{T}(E)$ is the transport distribution function (TDF), $\mu$ is the chemical potential,$f(E)$ is the Fermi-Dirac distribution function.
The TDF can be expressed with the following formula:
\begin{equation}
    \mathcal{T}(E) = v^2(E)\tau(E)\mathcal{D}(E),
\end{equation}
where $v$ is the velocity, $\tau$(E) is the relaxation time, and $\mathcal{D}(E)$ is the density of states (DOS). 

Under RTA, the relaxation time is inversely proportional to the DOS 
\begin{equation}\label{taudos}
    \tau(E)=\frac{\mathcal{C}}{\mathcal{D}(E)}
\end{equation}
where $\mathcal{C}$ is scattering coefficient. For 2D material, the longitudinal velocity $v(E)$ is related to the group velocity as written below
\begin{equation}
    v^2(E) = \frac{v^2_g}{2},
\end{equation}
where $v_g = \frac{1}{\hbar}\frac{dE}{dk}$ and $d$ is the dimension of the material. Then we get,

\begin{equation}\label{velocity}
    v^2(E) = \frac{n^2A^2_n}{2\hbar^2} \left(\frac{E-\Delta}{A_n}\right)^{\frac{2(n-1)}{n}}.
\end{equation}

For each electronic band, the transport coefficient will be calculated within proper integration boundaries. By using Boltzmann transport theory within relaxation time approximation the transport coefficients of the thermoelectric properties can be calculated ~\cite{hasdeo2019optimal}
\begin{equation}
    \sigma = q^2 \mathcal{L}_0,
\end{equation}
\begin{equation}\label{Seebeck}
    S = \frac{1}{qT} \frac{\mathcal{L}_1}{\mathcal{L}_0},
\end{equation}
and 
\begin{equation}
    \kappa_e = \frac{1}{T} \left(\mathcal{L}_2 - \frac{(\mathcal{L}_1)^2}{\mathcal{L}_0}\right).
\end{equation}
In the case of the multiband system, the TE properties can be calculated by summing up the contribution from all the bands involved: 
\begin{equation}
    \sigma = q^2 \sum_{j=1}^{N_B}\mathcal{L}_{0j},
\end{equation}
\begin{equation}
    S = \frac{1}{qT} \frac{\sum_{j=1}^{N_B}\mathcal{L}_{1j}}{\sum_{j=1}^n\mathcal{L}_{0j}},
\end{equation}
and 
\begin{equation}
    \kappa_e = \frac{1}{T} \left(\sum_{j=1}^{N_B}\mathcal{L}_{2j} - \frac{(\sum_{j=1}^{N_B}\mathcal{L}_{1j})^2}{\sum_{j=1}^{N_B}\mathcal{L}_{0j}}\right)
\end{equation}
where $i$ is the index of the band and $n$ is the number of bands which are considered.
From those parameters, the figure of merit can be calculated as
\begin{equation}
    ZT = \frac{S^2\sigma}{\kappa_e+\kappa_{ph}} T,
\end{equation}
where $\kappa_{ph}$ is the lattice thermal conductivity and the power factor, PF
\begin{equation}
    PF = S^2\sigma.
\end{equation}

The TDF of each band using approximation in Eqs.~\eqref{taudos} and~\eqref{velocity} can be written as
\begin{equation}
    \mathcal{T}^{(n)}(E) = \mathcal{C} \frac{n^2A^2_n}{2\hbar^2} \left(\frac{E-\Delta}{A_n}\right)^{\frac{2(n-1)}{n}}
\end{equation}
By using dimensionless variables $\varepsilon = E/k_BT$, $\eta = \mu/k_bT$,  and $\tilde{\Delta} = \Delta/k_BT$ then the TE integral for Dirac conduction band
\begin{equation}
    \mathcal{L}^{(1)}_{i,c}=\frac{A^2_1\mathcal{C}}{2\hbar^2}(k_BT)^i \mathcal{F}_{i,c}(\eta-\tilde{\Delta}) 
\end{equation}
with
\begin{equation}
    \mathcal{F}_{i,c}(\eta) = \int_{-\eta}^\infty \frac{x^i}{(e^x+1)^2} dx.
\end{equation}

For the Dirac valence band we get
\begin{equation}
        \mathcal{L}^{(1)}_{i,v}=\frac{A^2_1\mathcal{C}}{2\hbar^2}(k_BT)^i \mathcal{F}_{i,v}(\eta+\tilde{\Delta}) 
\end{equation}
with 
\begin{equation}
    \mathcal{F}_{i,v}(\eta) = \int_{-\infty}^{-\eta} \frac{x^i}{(e^x+1)^2} dx.
\end{equation}

For the parabolic, pudding mold 4th and 6th order, the TE integral can be written as
\begin{equation}
    \begin{split}
    \mathcal{L}^{(2)}_{i}=&\frac{4A_2\mathcal{C}}{2\hbar^2}(k_BT)^{i+1} \left[\mathcal{F}_{i+1}(\eta-\tilde{\Delta})+ (\eta-\tilde{\Delta})\right. \\ &\left. \times\mathcal{F}_i(\eta-\tilde{\Delta})-\mathcal{F}_{i+1}(\eta-\tilde{E}_0) \right.\\
    &\left.-(\eta-\tilde{\Delta})\mathcal{F}_i(\eta-\tilde{E}_0)\right],
\end{split}
\end{equation}

\begin{align}
    \mathcal{L}^{(4)}_{i}=\frac{16(A_4)^{\frac{1}{2}}\mathcal{C}}{2\hbar^2}(k_BT)^{i+\frac{3}{2}}\mathcal{H}_{4,i}(\eta,\tilde{\Delta},\tilde{E}_0)
\end{align}
with 
\begin{align}
    \mathcal{H}_{4,i}(\eta,\tilde{\Delta},\tilde{E}_0) = \int_{\tilde{\Delta}}^{\tilde{E}_0} (\varepsilon-\tilde{\Delta})^{\frac{3}{2}}(\varepsilon-\eta)^i\frac{e^{\varepsilon-\eta}}{(e^{\varepsilon-\eta}+1)^2} d\varepsilon,
\end{align}
\begin{align}
    \mathcal{L}^{(6)}_{i}=\frac{36(A_4)^{\frac{1}{3}}\mathcal{C}}{2\hbar^2}(k_BT)^{i+\frac{5}{3}} \mathcal{H}_{6,i}(\eta,\tilde{\Delta},\tilde{E}_0)
\end{align}
with
\begin{align}
    \mathcal{H}_{6,i}(\eta,\tilde{\Delta},\tilde{E}_0) = \int_{\tilde{\Delta}}^{\tilde{E}_0} (\varepsilon-\tilde{\Delta})^{\frac{5}{3}}(\varepsilon-\eta)^i\frac{e^{\varepsilon-\eta}}{(e^{\varepsilon-\eta}+1)^2} d\varepsilon,
\end{align}
respectively.

The units of thermoelectric properties $S_0$, $\sigma_0$, and $\kappa_0$ are given by
\begin{equation}
    S_0 = \frac{k_B}{e} = 86.17\, \mu \text{V/K},
\end{equation}
\begin{equation}
    \sigma_0 = \frac{e^2\mathcal{C}}{2 \hbar},
\end{equation}
and 
\begin{equation}
    \kappa_0 = \frac{1}{2 \hbar T}.
\end{equation}

\section{Results and Discussion}
\label{sec:res}

\subsection{Three-band systems}
\label{subsec:3band}
We calculate the thermoelectric properties of band structures which consist of three bands. Here we use $T = 300$K which corresponds to $k_BT = 0.026$ eV. The gap between Dirac bands is set to $2k_BT$ while the bandwidth of the nonlinear band is $5$ eV. We assume the ideal system which the lattice thermal conductivity $\kappa_{ph} =0$. 

\begin{figure}[ht!]
    \centering
    \includegraphics[width=8cm]{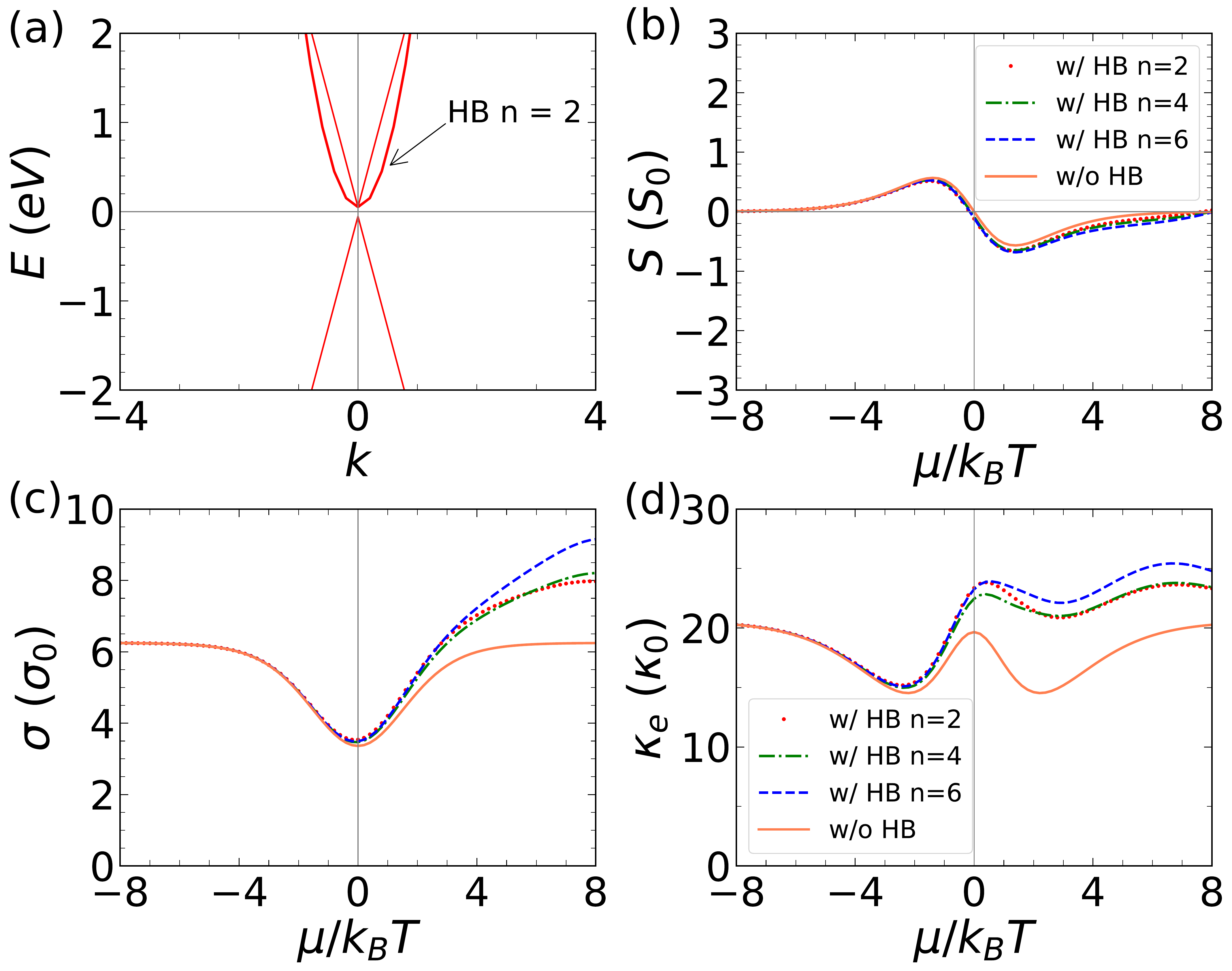}
    \caption{Thermoelectric properties of a three-band system with symmetric Dirac bands. The band structure in (a), the Seebeck coefficient ($S$) in (b), electrical conductivity in (c), and electronic thermal conductivity in (d) for each combination. The results for $S,\sigma$ and $\kappa_e$ are expressed in the units of $S_0$, $\sigma_0$, and $\kappa_0$ respectively.}
    \label{3combined-symmetry}
\end{figure}

First, we calculate the thermoelectric properties of three-band model with these set of constants $A_1^c = 2.5\, \text{eVm}$ , $A_1^v = -2.5\, \text{eVm}$, $A_2 = 2.5\, \text{eVm}^2$, $A_4 = 2.5\, \text{eVm}^4$ and $A_6 = 2.5 \, \text{eVm}^6$. We plot the band structure model for a three-band system with parabolic inside in Fig.~\ref{3combined-symmetry}(a) and its thermoelectric properties such as Seebeck coefficient, electrical conductivity, and electronic thermal conductivity in Figs.~\ref{3combined-symmetry}(b-d). We plot the thermoelectric properties as a function of a dimensionless variable from its chemical potential $\mu$. For each thermoelectric property, we plot each type of three-band system with a different heavy band. In addition to the three-band system, we also plot the thermoelectric properties of the pure Dirac material system.

In Fig.~\ref{3combined-symmetry}(b), we have plotted the Seebeck coefficient for those three-band systems. First, we can see that the value of the Seebeck coefficient is similar to all of the systems in negative chemical potential as shown in that the curves coincide. Because in this negative region, only the Dirac band exists for all of the systems. At the Fermi level ($\mu/k_BT = 0$), it starts to have a different value for each heavy band system. The difference between each system becomes more evident in the positive chemical potential. The value of the Seebeck coefficient of the three-band systems is just slightly larger than the Dirac cones system. It is due to the existence of a heavy band in the system that generates energy filtering for impeding the low carrier to get excited. This process will imbalance the number of careers available in the band. As we know that the more imbalance carrier between the band, the higher the Seebeck coefficient will be.

In the case of electrical conductivity in Fig.~\ref{3combined-symmetry}(c), the electrical conductivity in the vicinity of the Fermi level for the three-band system is slightly higher than the Dirac material. It becomes much different in the positive chemical potential region. For the Dirac material, the electrical conductivity is symmetric with respect to the Fermi level (between p-type and n-type) while the three-band system is not symmetric anymore. The highest electrical conductivity is achieved by the pudding-mold 6th order system in positive chemical potential. This large value comes from the dependence of $n$ in the thermoelectric integral $\mathcal{L}_0$ wherein the Seebeck coefficient cancels out because of the division of $\mathcal{L}_1$ and $\mathcal{L}_0$ in Eq.~\eqref{Seebeck}. On the other hand, the more evident difference among the nonlinear band appears also in the electronic thermal conductivity. The electronic thermal conductivity is somewhat more related to the electrical conductivity than the other TE properties. Near the Fermi level, the electrical conductivity of parabolic and pudding-mold 6th order is similar which is also higher than the pudding-mold 4th order. In the positive chemical potential, the electrical conductivity of the pudding-mold 6th order system becomes larger. This suggests that the three-band system will provide better TE performance on n-doping than p-doping.

\begin{figure*}[ht!]
    \centering
    \includegraphics[width=17cm]{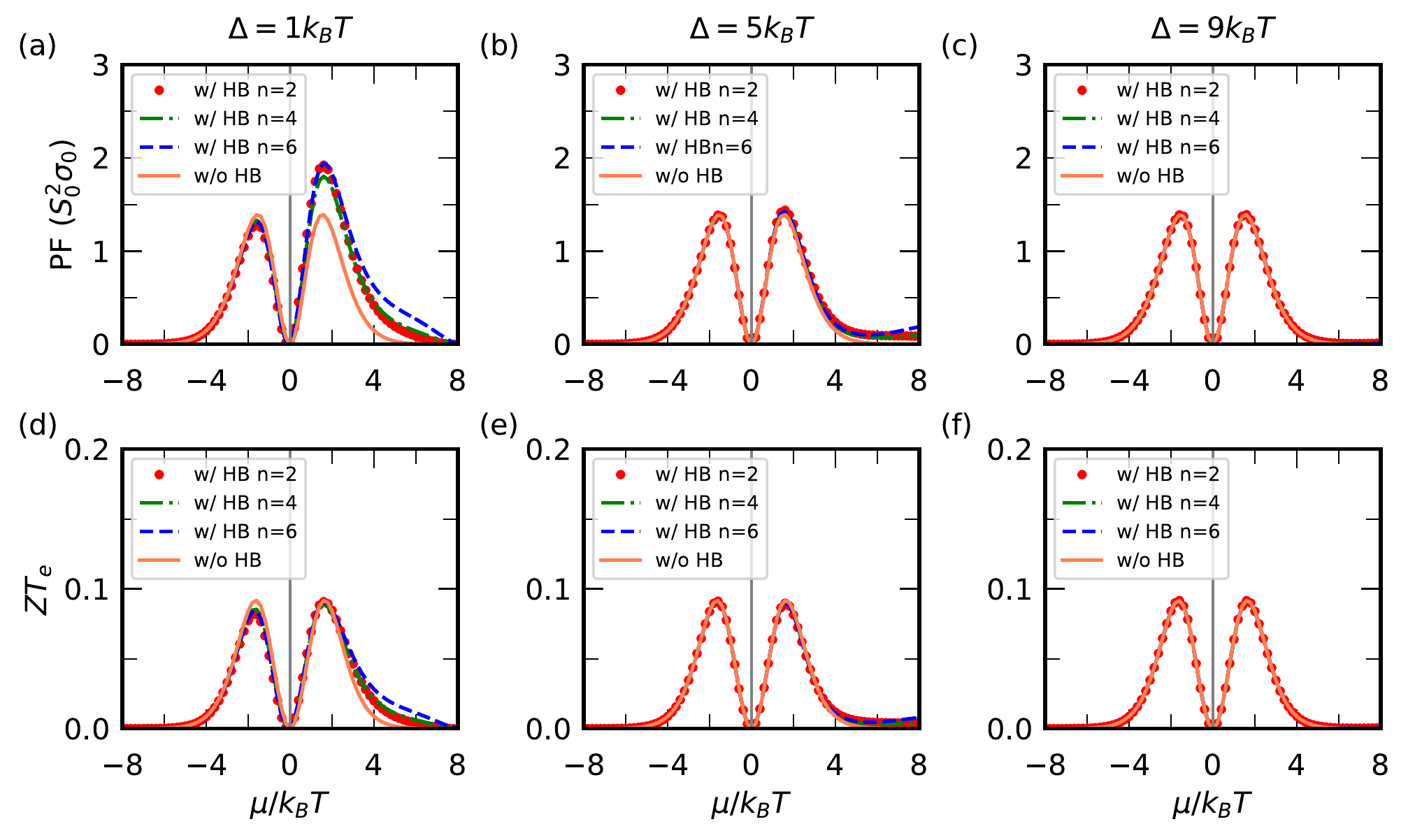}
    \caption{Power factor (PF) in (a-c) and electronic figure of merit ($ZT_e$) in (d-f) within three-band model. The position of the nonlinear band is varied corresponding to the Fermi level. The nonlinear band is shifted up $1k_BT$ in (a) and (d), $5k_BT$ in (b) and (e), and $9k_BT$ in (c) and (f).}
    \label{PFZT3-symmetry}
\end{figure*}

In general, the Seebeck coefficient will be higher if the career concentration is small enough and there is a difference between the conduction band career and valence band career that are electron and hole respectively. The Seebeck coefficient for each nonlinear three-band system is rather similar. We see many differences between this nonlinear band system in the electrical conductivity and electronic thermal conductivity. From these three TE properties, we plot the power factor and the figure of merit in Figs.~\ref{PFZT3-symmetry}. We did some changes to the position of the nonlinear band. We shifted up $4k_BT$ and $8k_BT$ in Figs.\ref{PFZT3-symmetry}(b and e) and (c and f) respectively. We could see a higher PF when the nonlinear band intersects the Dirac band at $\Delta = 1k_BT$. As the nonlinear band shifted up, the value of PF of the three-band system will reduce to the PF of Dirac material. We could see that the combination with pudding-mold 6th order features the highest PF among the others in the positive region of chemical potential.

However, from Figs.~\ref{PFZT3-symmetry}(c-f) we can see that the figure of merit of the three-band system is much similar to the Dirac material. The most evident difference can be seen when the nonlinear band intersects the same point with the Dirac band compared to others ($\Delta = 5k_BT$ and $\Delta = 9k_BT$). Despite higher PF for the three-band system where the nonlinear band intersects the Dirac band, the large electronic thermal conductivity $\kappa_e$ of this system cancels the promising value of PF as the figure of merit is inversely proportional to $\kappa_e$. We also noted that the further we shift up the nonlinear band, it will reduce to a two-band system as if the nonlinear band does not exist and only Dirac cones exist in the system.

One example of real material that represents the most similar to this three-band model was found on the square-octagonal \ch{MoS2}. From the band structure in Fig. \ref{so-MoS2BandDOS}, we can see that there are Dirac cone and a parabolic band near the Fermi level. By doing numerical fitting to this DFT band structure using the dispersion relation in Eq.~\eqref{eq:disp}, we found $A_1^c = 2.472 \, \text{eVm}$, $A_1^v = -2.556 \, \text{eVm}$, and $A_2 = 2.563 \, \text{eVm}^2$. Unfortunately, the thermoelectric properties in Fig.~\ref{3combined-symmetry} are not in the same trend as shown from the first principle result in Fig.~\ref{TEBoltzTrap2} though we have used a similar value of the coefficient $A_n$ for each respective band. 

\begin{figure}[ht!]
    \centering
    \includegraphics[width=8cm]{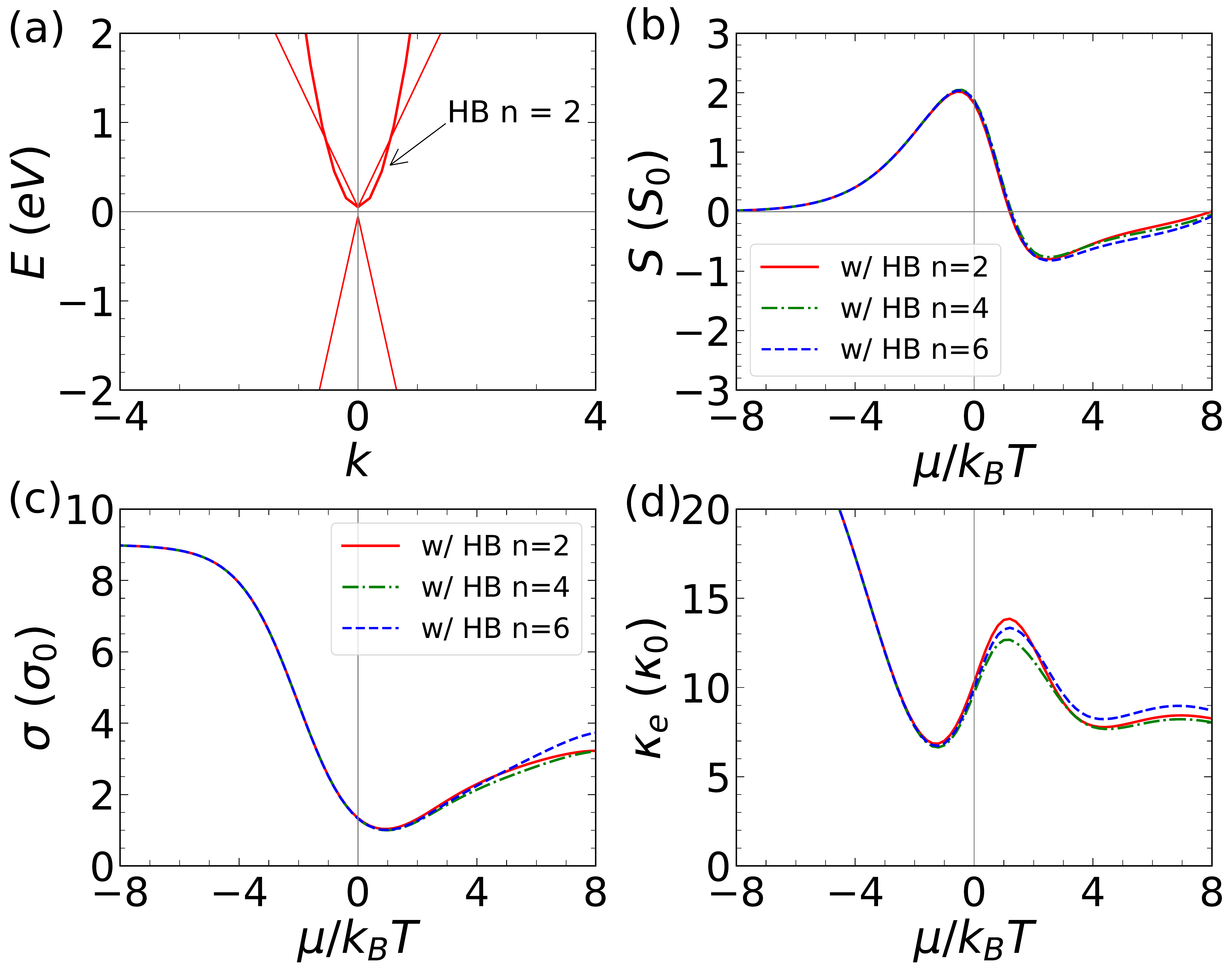}
    \caption{Thermoelectric properties of a three-band system with asymmetric Dirac bands. The band structure in (a), the Seebeck coefficient ($S$) in (b), electrical conductivity in (c), and electronic thermal conductivity in (d) for each combination. The results for $S,\sigma$ and $\kappa_e$ are expressed in the units of $S_0$, $\sigma_0$, and $\kappa_0$ respectively.}
    \label{3combined-assymmetry}
\end{figure}

From the previous result, we do some changes deliberately to these sets of coefficients to get a similar result to the first principle calculation. We found these set of coefficients $A_1^c = 1.4 \, \text{eVm}$, $A_1^v = -3.3 \, \text{eVm}$, and $A_2 = 2.5 \, \text{eVm}^2$ give the similar result to the first principle result as shown in Fig~\ref{3combined-assymmetry}. Compared to the numerical fitting result, this set of coefficients is much different from the linear band coefficient. This difference can be understood as an effective model to capture the effect of the other bands. In the first principle calculation, there is more than three bands are considered to generate the thermoelectric properties of the material.

\subsection{Two-band systems}
\label{subsec:2band}
\begin{figure}[ht!]
    \centering
    \includegraphics[width=8cm]{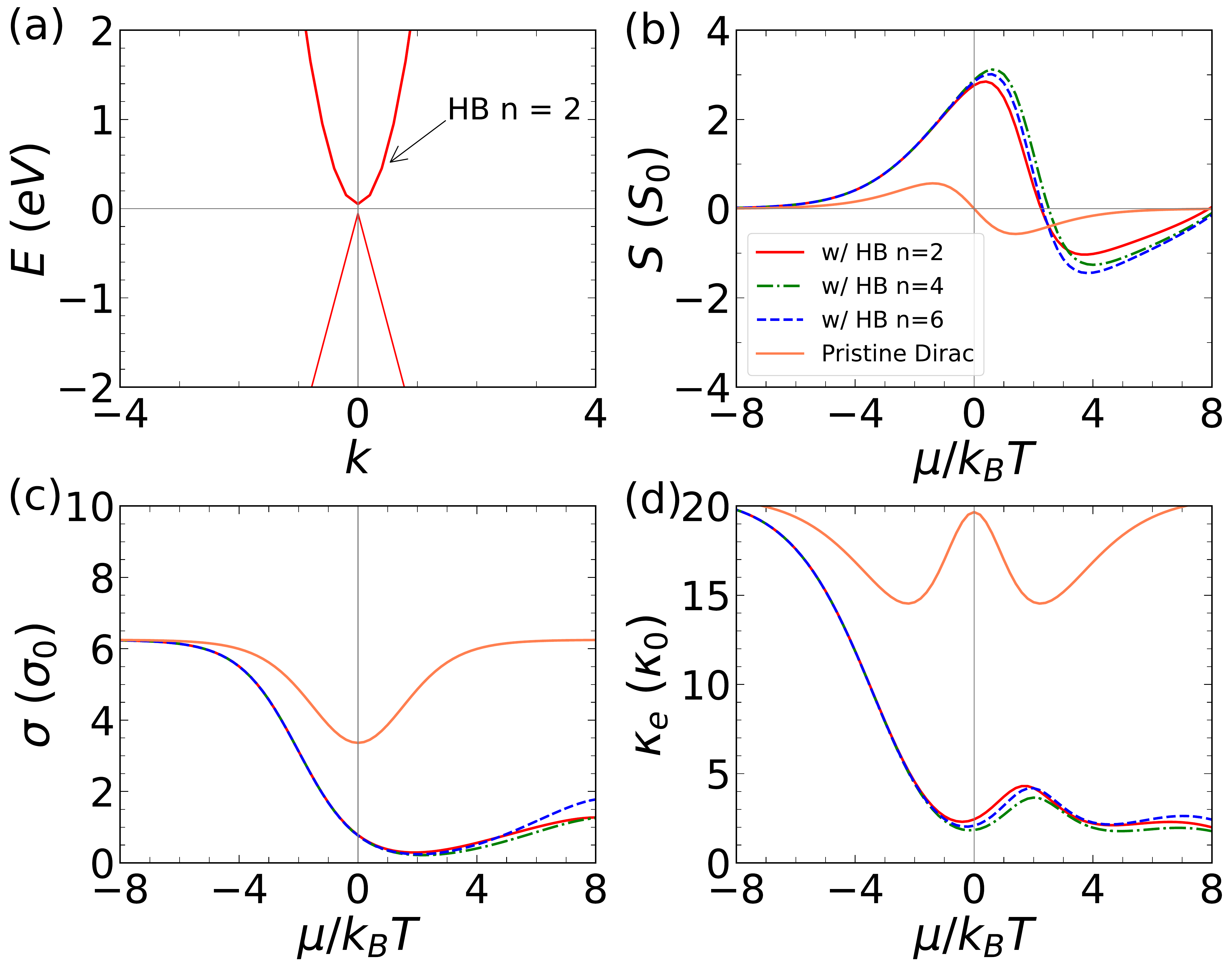}
    \caption{Thermoelectric properties of the two-band system. Seebeck coefficient ($S$) in (a), electrical conductivity in (b), and electronic thermal conductivity in (c) for each combination. The results for $S,\sigma$ and $\kappa_e$ are expressed in the units of $S_0$, $\sigma_0$, and $\kappa_0$ respectively.}
    \label{2combined}
\end{figure}

Now, we remove the linear conduction band to make a two-band system. The two-band system is similar to the Dirac material system unless the conduction linear band is substituted by a nonlinear band as we can see in Fig.~\ref{2combined}(a). Doing the same calculation for the thermoelectric properties, we can see from Fig.~\ref{2combined}(b) that the Seebeck coefficient is generally higher than the three-band system or even the pristine Dirac material system. In this two-band system, we can see from Fig. \ref{2combined}(b) that the difference for each combination is slightly more evident than in the three-band systems. The two-band model that consists of pudding-mold 4th order features a higher Seebeck coefficient though it is just a tiny difference. This large Seebeck coefficient comes from the asymmetry condition in this two-band system. In the case of a pristine Dirac material system, we plot the same Dirac or linear band in both the valence and conduction band. This will generate a small Seebeck coefficient.

Moreover, we can see from Fig. \ref{2combined}(c) that the electrical conductivity is lower than in the three-band system for all combinations. Lower electrical conductivity will also compensate for the higher Seebeck coefficient that we got previously. Consequently, the power factor will not generally be higher than in the three-band system. Figure \ref{2combined}(f) depicts the electron thermal conductivity for the two-band system. The electron thermal conductivity is lower than in the three-band systems. It will result in delivering a higher power factor and $ZT_e$.

\begin{figure*}[ht!]
    \centering
    \includegraphics[width=17cm]{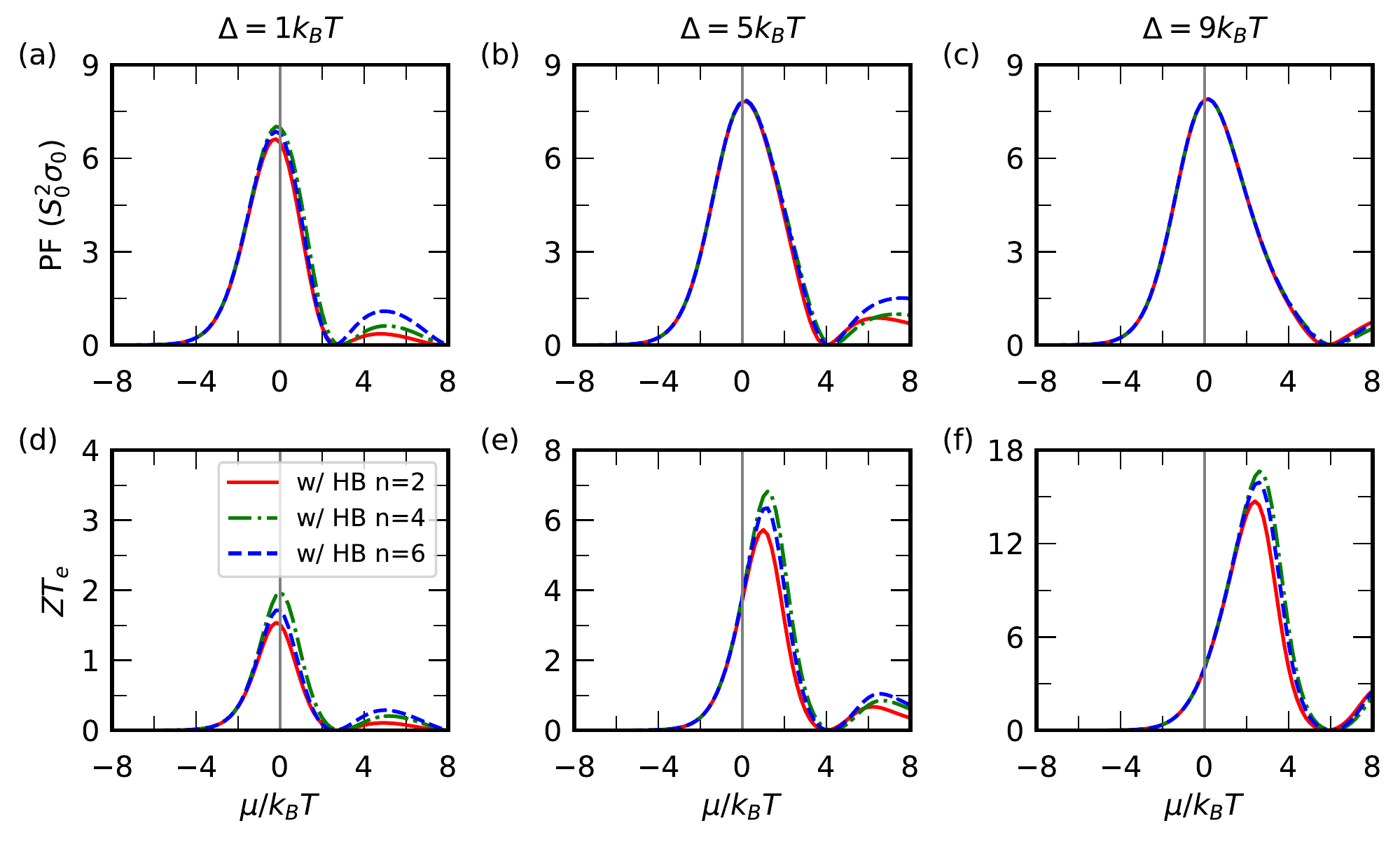}
    \caption{Power factor (PF) in (a-c) and electronic figure of merit ($ZT_e$) in (d-f) within two-band model. The position of the nonlinear band is varied corresponding to the Fermi level. The nonlinear band is shifted up $1k_BT$ in (a) and (d), $5k_BT$ in (b) and (e), and $9k_BT$ in (c) and (f).}
    \label{PFZT2}
\end{figure*}

Now we shift up the nonlinear band from the original position $1k_BT$ to some higher values such as $5k_BT$ and $9k_BT$. From Figs.~\ref{PFZT2}(a-c), the power factor will increase as the nonlinear band is shifted up. However, the power factor is similar between $\Delta 5k_BT$ and $\Delta 9k_BT$. The large difference between the two-band and the three-band system comes in the value of the electronic figure of merit $ZT_e$ in Figs.~\ref{PFZT2}(d-f). The more we shifted up the nonlinear band, means higher the bandgap, the higher $ZT_e$ will be. The smallest $ZT_e$ when $\Delta = 1k_BT$ for a two-band system is 10 times larger than the same $\delta$ for a three-band system. For the larger bandgap in Figs.\ref{PFZT2}(e and f), the $ZT_e$ becomes extremely larger than the previous one. Another fact that we found in this two-band system is the clear difference between parabolic, pudding-mold 4th, and pudding-mold 6th band in the $ZT_e$. We can see from Figs.~\ref{PFZT2}(d-f) that the two-band system with a pudding-mold band is higher than the parabolic system as it is presumed from the theory that the pudding-mold band has a drastic rise in the DOS as shown in Fig.~\ref{bandDOS2}(d and f).

\subsection{Effect of bandwidth}
\begin{figure*}[ht!]
    \centering
    \includegraphics[width=17cm]{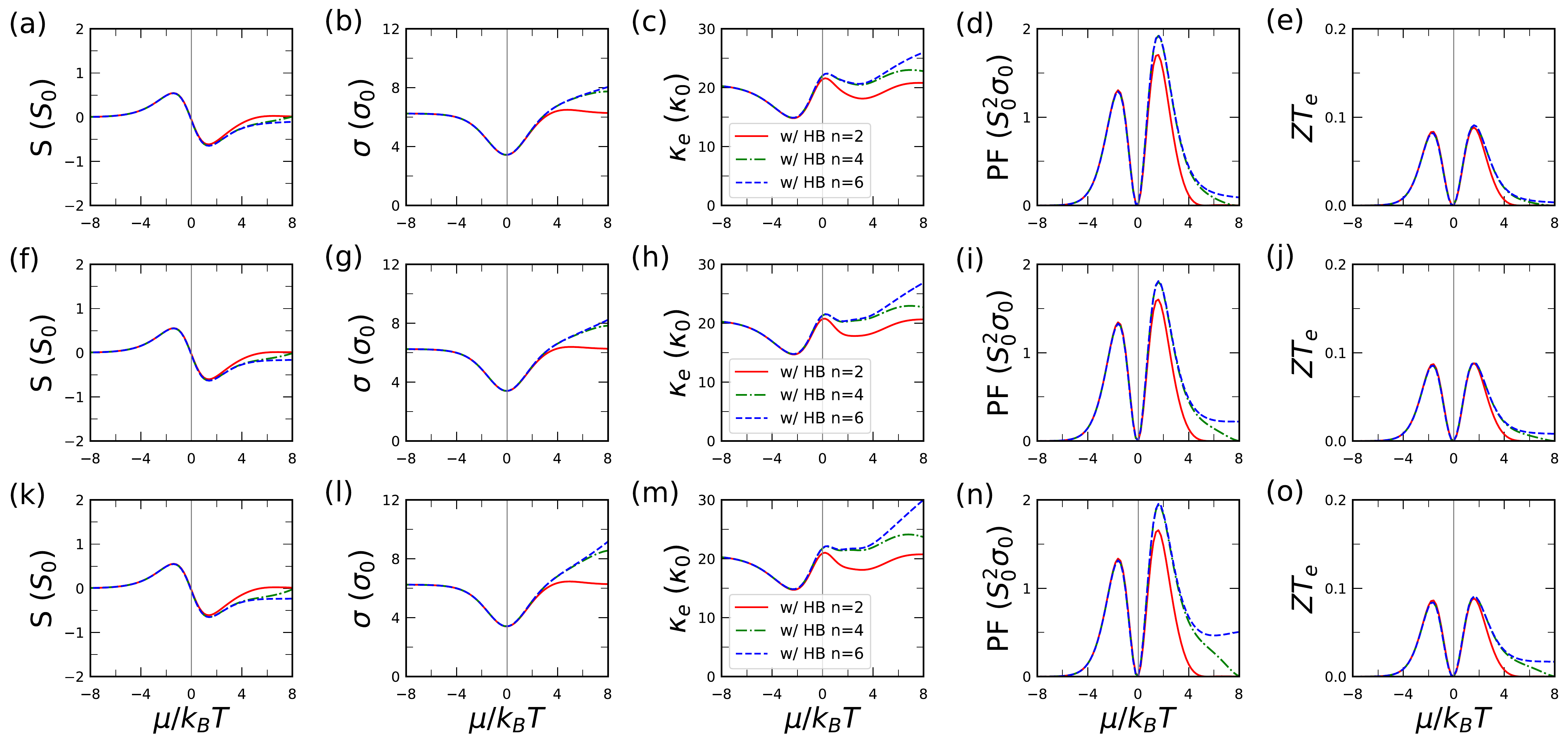}
    \caption{Seebeck coefficient, electrical conductivity, electronic thermal conductivity, power factor, and $ZT_e$ within three-band models with parabolic in (a-e), pudding-mold 4th in (f-j), and pudding mold 6th band in (k-o) varied by different bandwidth of the nonlinear band.}
    \label{3band-bandwidth}
\end{figure*}

Here, we will present the effect of the bandwidth of the nonlinear band on the thermoelectric properties. A previous study point out that there was no optimum bandwidth for $\tau = \frac{\mathcal{C}}{DOS}$ scattering approximation\cite{PhysRevLett.107.226601}. In a three-band system, we found that there exists an optimum value of the bandwidth to bring the optimum thermoelectric properties. The Seebeck coefficient in Fig. ~\ref{3band-bandwidth} (a,f,k), there is no large difference for each bandwidth in every type of nonlinear band. In contrast with the Seebeck coefficient, the electrical conductivity (Fig.~\ref{3band-bandwidth} (b,g,l) and electronic thermal conductivity (Fig.~\ref{3band-bandwidth} (c,h,m)) do have apparent difference for each bandwidth, specifically in the positive regime of the Fermi level. Less bandwidth utilized features less electrical conductivity and electronic thermal conductivity. This results in a higher power factor for the higher bandwidth. From Fig.~\ref{3band-bandwidth}(d, i,n), we can see that the largest PF is achieved when the bandwidth is set to $10k_BT$. Increasing the bandwidth above $10k_BT$ such as $15k_BT$ does not give us a higher PF than the former one. Lastly, the figure of merit in Fig.~\ref{3band-bandwidth} (e,j,o) show a similar result for each bandwidth. The large power factor is compensated by the large electronic thermal conductivity. It results in a lower figure of merit (less than $0.1$).

\begin{figure*}[ht!]
    \centering
    \includegraphics[width=17cm]{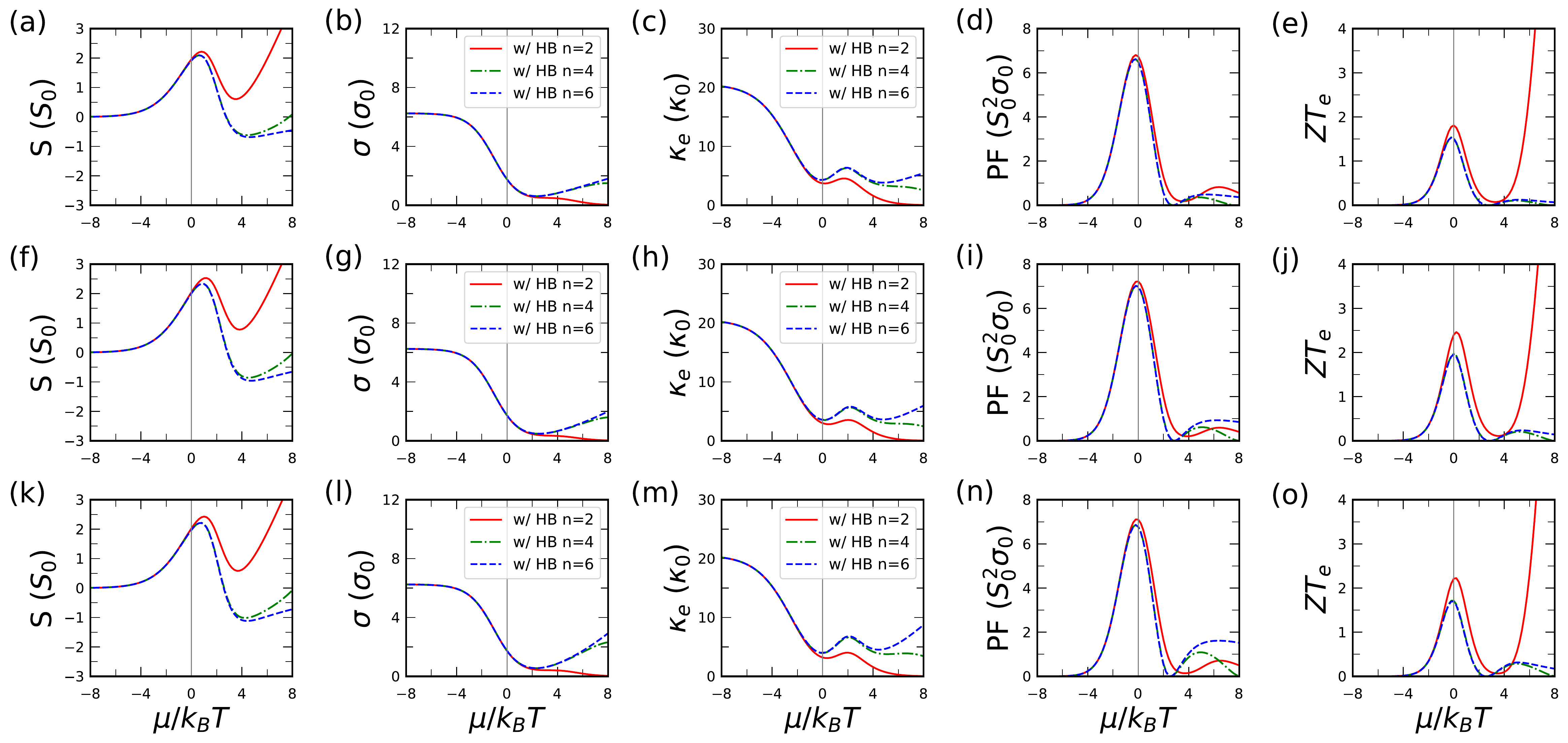}
    \caption{Seebeck coefficient, electrical conductivity, electronic thermal conductivity, power factor, and $ZT_e$ within two-band models with parabolic in (a-e), pudding-mold 4th in (f-j), and pudding mold 6th band in (k-o) varied by different bandwidth of the nonlinear band.}
    \label{2band-bandwidth}
\end{figure*}

We apply the same calculation for the two-band system. In this case, we found that the Seebeck coefficient for small bandwidth ($5k_BT$) in Fig.~\ref{2band-bandwidth}(a,f,k) is higher than the others at the chemical potential above the Fermi level. A similar trend to the three-band system, the electrical conductivity (Fig.~\ref{2band-bandwidth}(b,g,l)) and the electron thermal conductivity (Fig.`\ref{PFZT2}(c,h,m) is less in lower bandwidth. However, the power factor (Fig.~\ref{2band-bandwidth}) for each bandwidth does not have any large difference among them. The clear difference between the bandwidth happens in the figure of merit (Fig.~\ref{2band-bandwidth}(e,j,o)) where the small bandwidth features larger $ZT_e$ than the higher bandwidth.  It is physically reasonable because, in a two-band system, less bandwidth of the nonlinear band will make this band looks flatter.

\section{Conclusions}
\label{sec:con}
We find that in the three-band systems, any nonlinear band accompanying the Dirac band will generate similar TE properties near the Fermi level. The difference will emerge mostly in the positive chemical potential of the pudding-mold band system, which features a higher power factor and electronic figure of merit. The optimum thermoelectric power factor is found if the nonlinear band intersects the Dirac conduction band. The three-band system will reduce to a two-band Dirac material system when the nonlinear band is more than $k_BT$ above the Fermi level. The two-band system exhibits better TE performance in general than the three-band system. Compared to the three-band system, the difference given by the various nonlinear bands becomes more obvious in the two-band system, where the pudding-mold system offers the higher electronic figure of merit ($ZT_E$). We can conclude that fewer bands will give better TE performance than multiband systems but a single-band thermoelectrics in reality is not feasible. Therefore, materials that exhibit two bands near the Fermi level will give the best thermoelectric performance. 

\begin{acknowledgements}
A.D. performed first principles calculations using King Abdulaziz University (KAU)’s High Performance Computing Center (Aziz Supercomputer). He is also supported by the KAU scholarship. E.S. and A.R.T.N acknowledge computational facilities from BRIN HPC.
\end{acknowledgements}

\appendix

\section{Thermoelectrics of monolayer so-MoS$_2$}

\begin{figure}[ht!]
    \centering
    \includegraphics[width=8cm]{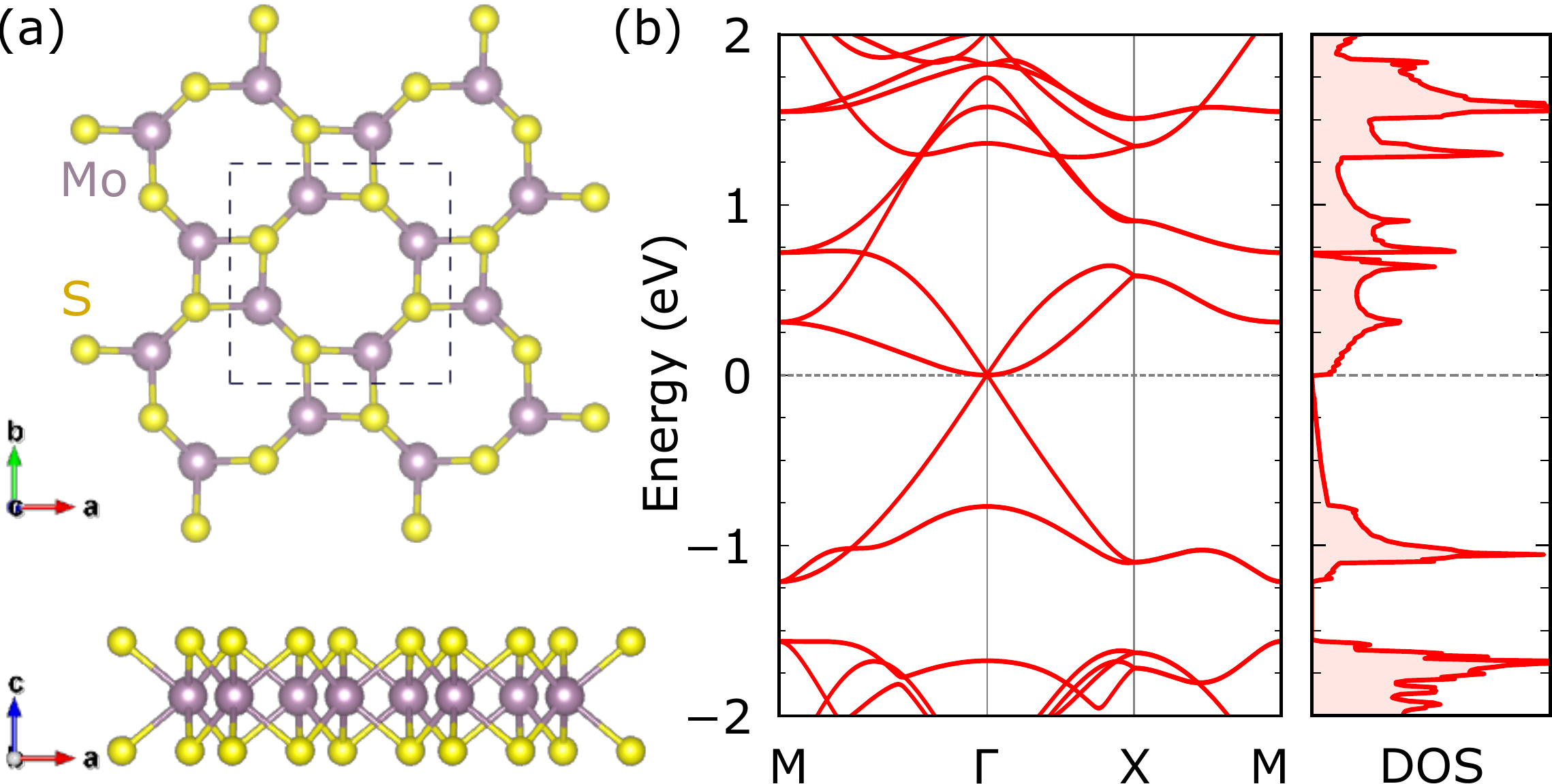}
    \caption{(a) The top and side view of so-\ch{MoS2}. (b) The electronic band structures and density of states.}
    \label{so-MoS2BandDOS}
\end{figure}

\begin{figure}[ht!]
    \centering
    \includegraphics[width=8cm]{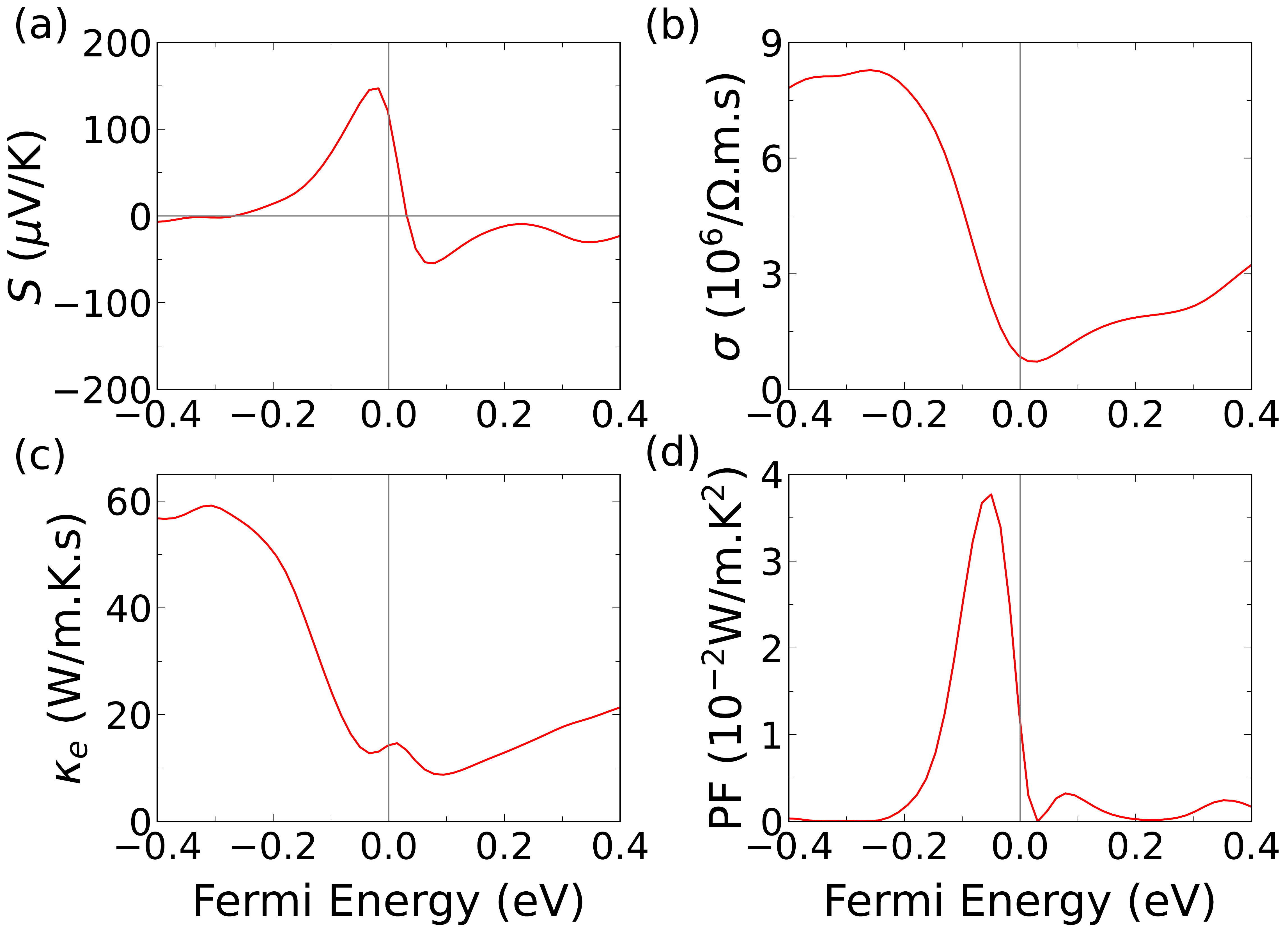}
    \caption{Thermoelectric properties of monolayer so-\ch{MoS2}: (a) Seebeck coefficient, (b) Electrical conductivity, (c) electronic thermal conductivity, (d) power factor}
    \label{TEBoltzTrap2}
\end{figure}

Here, we present the first principle calculation of monolayer so-\ch{MoS2} which has three bands near the fermi level. The so-\ch{MoS2} is constructed of repeating square-octagon pairs in a square-lattice, as shown in Fig.~\ref{so-MoS2BandDOS}(a). It has a parabolic band and Dirac bands, as shown in Fig.~\ref{so-MoS2BandDOS}(b). The band structure calculation is performed by~{\sc{Quantum ESPRESSO}}~\cite{QE,giannozzi2009quantum} within the optimized norm-conserving Vanderbilt (ONCV) pseudopotentials~\cite{hamann2013, schlipf2015} and the Perdew-Burke-Ernzerhof~\cite{PBE} parameterization for the exchange-correlation functional. We set vacuum layer of 25~\AA~to avoid the interlayer interactions due to lattice periodicity. The plane-wave basis sets with a kinetic energy cutoff of 100 Ry were also employed in the calculation. We sample the Brillouin zone using $20 \times 20 \times 1$ Monkhorst-Pack grids~\cite{tetrahedron} for the density of states (DOS) calculation. Then, we scale the DOS by the imaginary part of the electron self-energy from previous calculation~\cite{xia2019leveraging} to count the relaxation time by the electron-phonon interaction given by $\tau(E) = C/\mathrm{DOS}(E)$. For monolayer so-\ch{MoS2}, the fitting parameter is set to $C=3.29\times10^{-14}~s/eV$. In Figure~\ref{TEBoltzTrap2}, we plot the thermoelectric (TE) transport coefficients of monolayer so-\ch{MoS2} at $T=300$ K. The TE properties are calculated from Boltzmann transport equation under the relaxation time approximation, as implemented in {\sc{BoltzTraP2}} code~\cite{boltztrap2}.

\bibliography{references}

\end{document}